# Superconducting-superconducting hybridization for enhancing single-photon detection

Yachin Ivry,[1,2,3,*] Jonathan J. Surick,[3] Maya Barzilay,[1,2] Chung-Soo Kim,[3] Faraz Najafi,[3] Estelle Kalfon-Cohen,[4] Andrew D. Dane[3] and Karl K. Berggren[3,*]

*Correspondence to: ivry@technion.ac.il and berggren@mit.edu.

1. Department of Materials Science & Engineering, Technion – Israel Institute of Technology, Haifa, 32000, Israel

2. Solid State Institute, Technion – Israel Institute of Technology, Haifa, 32000, Israel.

3. Research Laboratory of Electronics, Massachusetts Institute of Technology, 77 Massachusetts Avenue, Cambridge, Massachusetts 02139, USA

4. Department of Aeronautics and Astronautics, Massachusetts Institute of Technology, Cambridge, Massachusetts 02139, USA

Abstract

The lack of energy dissipation and abrupt electrical phase transition of superconductors favorite them for nanoscale technologies, including radiation detectors, and quantum technologies. Moreover, understanding the nanoscale behavior of superconductivity is significant for revealing the onset of collective-electron behavior in nature. Nevertheless, the limited number of accessible superconductors restricts availability of the superconducting properties, encumbering the realization of their potential. Superconducting nanowire single photon detectors (SNSPDs) sense single-IR photons faster and more efficient with respect to competing technologies. However, these advantageous properties are material-dependent causing an undesirable speed-efficiency payoff. Usually, SNSPDs based on granular materials are faster, while those based on amorphous materials are more efficient. Here we optimized ultrathin films of granular NbN on $SiO_2$ and of amorphous $\alpha W_5Si_3$. We showed that hybrid SNSPDs made of 2-nm-thick $\alpha W_5Si_3$ films over 2-nm-thick NbN films exhibit advantageous coexistence of timing (< 5-ns reset time and 52-ps timing jitter) and efficiency (> 96% quantum efficiency) performance. We propose that the governing mechanism of this hybridization is the presence of a dual superconducting behavior: native superconductivity of each of the films and superconductivity that is induced from the neighboring film via the proximity effect. Our results not only demonstrate improvement in SNSPDs performance, but they also suggest that such hybridization can significantly expand the range of available superconducting properties, impacting other nano-superconducting technologies. Lastly, this hybridization may be used to tune properties, such as the amorphous character of superconducting films and to illuminate the elusive onset of collective-electron behavior near the superconducting-to-insulating transition.

Main text:

Detecting single photons at infrared wavelengths is central to next-generation communication and sensing technologies and a key requirement of quantum-cryptography and computing. Superconducting nanowire single-photon detectors (SNSPDs)[1] are capable of detecting photons at the near IR with: > 90% system efficiency,[2] timing jitter less than 30ps,[3] reset time of just a few nanoseconds[4] and with lower dark-count rate[2] than competing technologies.[5] For example, the recent loop-hole free demonstration of violation of Bell's inequality was made possible the availability of these detectors.[6,7] Similarly, recent demonstrations of ultrafast space-based optical communications relied on these detectors.[8] Future progress in the domain of quantum and classical communication and information processing thus requires continued reduction in jitter and reset time, and further increases in efficiency of this technology.

However, existing SNSPD materials have not been optimized to allow all of these advantages occurring simultaneously, *e.g.*, there is a fundamental tradeoff between speed and efficiency—detectors with high efficiency tend to require larger areas, and thus have longer reset times and longer jitter.[9] Optimization of this tradespace is hindered by the existence of a relatively limited number of superconducting materials that are available for use in SNSPDs, and the large number of detector parameters that have to be optimized, including: operating temperature, optical absorption at or across a variety of wavelengths, fabrication yield, material stability under thermal cycling, output signal amplitude, and resistance to dark counts. Tungsten silicide ($W_xSi_{1-x}$), niobium nitride (NbN), niobium titanium nitride (NbTiN), magnesium diboride, yttrium barium copper oxide, molybdenum germanium, molybdenum silicide, and niobium have all been used as superconducting nanowire detectors, although currently NbN, NbTiN, and $W_xSi_{1-x}$ based SNSPDs are preferred, having demonstrated detection efficiencies approaching the quantum limit (reflected by a saturating behavior of the detection efficiency as a function of bias current applied on these devices). Systems that include such devices have exhibited over 90% total system efficiency with dark count rates as low as < 1/sec.[2] However, such $W_xSi_{1-x}$ -based detectors have a number of drawbacks, including ~100 ps timing jitter, ~40 ns reset time, and they require lower operating temperatures (< 1 K and can be as low as 120 mK).[2] In these ways, they are inferior to the faster NbN detectors, which in turn demonstrate typical < ~35 ps timing jitter and ~3 ns reset time for similar device designs.[10] This situation invites us to consider modifying materials, to find routes to somehow engineer superior properties in a single device.

Previous attempts to externally tune superconducting properties of sensitive photodetectors include mainly the utilization of proximity effects between a superconductor and a normal metal.[11] That is, when a normal metal is introduced in close proximity to a superconductor, the normal electrons can diffuse into the superconductor, suppressing the superconducting transition temperature $T_C$. This $T_C$-tuning mechanism allowed *e.g.* in transition edge sensor (TES), deterministic tuning of device properties, such as the detected photon wavelength[12,13] (we should note that in principle, a counter effect where superconducting electrons are diffused into the normal metal is also possible, but is less common in superconducting-based photodetection technologies). An obvious approach to tuning the materials of NbN would then be to deposit a thin layer of normal layer on top of it. However, given that we know WSi has advantageous detector properties that are somewhat complementary to those of NbN, we decided to attempt to proximitize the NbN with WSi (*i.e.* another superconductor), rather than with a normal metal.

**Methods – fabricating hybrid materials and devices**

Because both $W_xSi_{1-x}$ and NbN span all the required properties, but neither is ideal on its own, we chose to fabricate the devices out of a $W_xSi_{1-x}$ - NbN bilayer in the hopes of better optimizing device performance. In order to understand the process better, we first demonstrated the feasibility of an ultrathin NbN superconductor, which we accomplished by growing 2nm-thick NbN films on $SiO_2$ substrates. We then demonstrated the feasibility of ultrathin $W_xSi_{1-x}$ superconductors, which we accomplished by growing and analyzing a variety of < 10-nm-thick $W_xSi_{1-x}$ films. After optimizing the growth of each of the materials separately, we fabricated and characterized a film stack of 2 nm NbN and 2 nm $W_xSi_{1-x}$. Finally, we fabricated and tested devices based on this new hybridized material.

The growth of NbN on $SiO_2$ films was optimized based on the initial parameters we previously reported for the thin NbN films on MgO.[14] We chose a thermally-oxidized $SiO_2$ layer (255nm thick) of a Si substrate over MgO or sapphire because of three main reasons (our typical substrate size was 1cm × 1cm). Firstly, under radiative heating, Si is expected to absorb more energy than MgO, allowing for higher in-situ substrate temperature, which in turn is known to help the growth of NbN.[15] Secondly, NbN has a good lattice matching with MgO, enabling a more epitaxial-like growth than on the amorphous $SiO_2$. Epitaxial growth may be advantageous when one is interested in examining material properties, but the long-range order that accompanies this growth might become a disadvantage when one aims at processing robust and reproducible devices whose orientation varies randomly with respect to the crystallographic axes. Lastly, because Si wafers can be thermally oxidized on both sides, one can use them in an optical stack that enhances the detection system efficiency.[16]

Our NbN on $SiO_2$ exhibited relatively-high $T_c$ values with respect to its sheet resistance at the normal state and film thickness (*e.g.* 8.3 K that was measured at the 90% drop of the resistance for $R_s = 1111.3$ Ω/□ and $d = 2$ nm). Our films were found to be of a continuous polycrystalline structure with a typical grain size comparable to the film thickness (Figure 1a). The width of the transition and the residual resistance ratio between the room-temperature $R_s$ and the sheet resistance just above the transition were $\Delta T_C$ = 1.6 K, RRR = 1. Figure 1b shows a typical superconducting transition curve of a $d$ = 2 nm film.

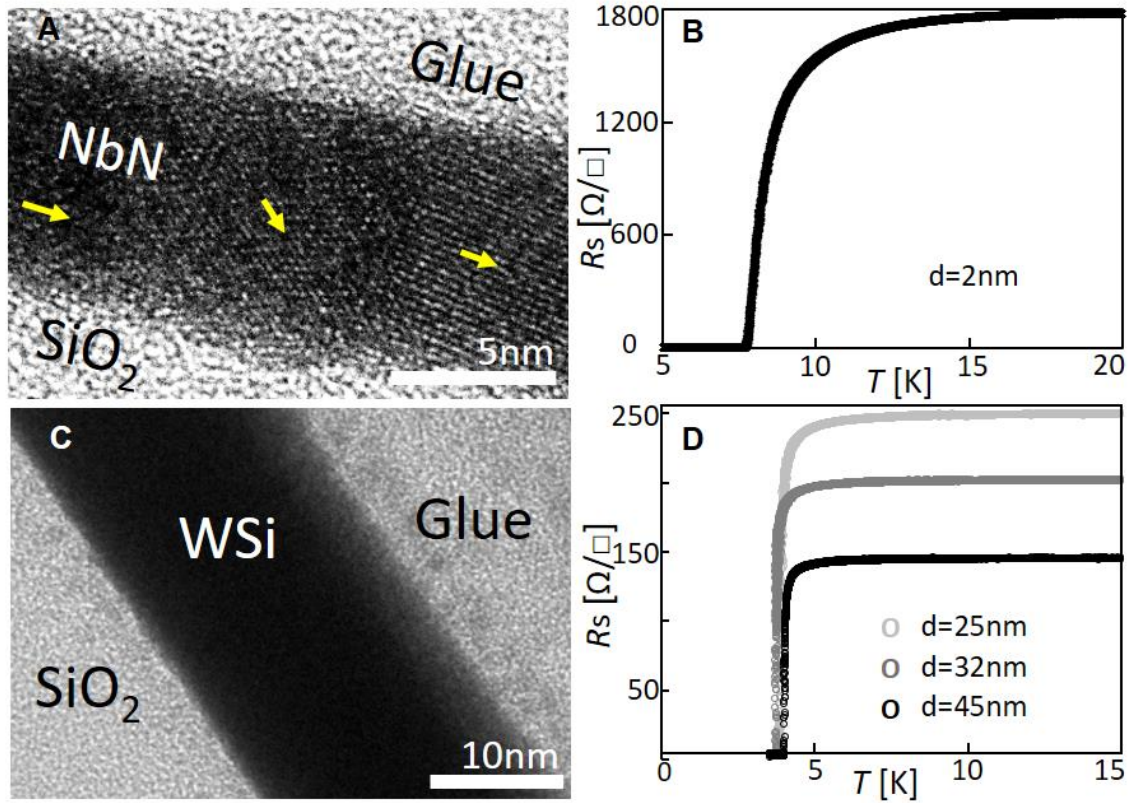

**Figure 1| Crystalline and amorphous thin superconducting films**. (**A**) A cross-section TEM micrograph of a 6.7-nm-thick NbN film on $SiO_2$ demonstrating its crystallinity and granularity (arrows highlight the different orientation of the grains). (**B**) The cooling curve of a 2-nm-thick film of an NbN thin film demonstrating a smooth transition and constant (within +/- 5%) resistance above $T_C$ (= 8.3 K), for such a thin film with $R_s$ = 1111.3 Ω/□, indicating the high quality of the superconductor. (**C**) A cross-section TEM micrograph of a thin $W_5Si_3$ film demonstrates its amorphous character. (**D**) The cooling curve of amorphous $W_5Si_3$ films demonstrate the same high-quality of the superconductor as the NbN. Here $T_C$ = 3.9 K for the $R_s$ = 239.7 Ω/□ film.

The superconducting behavior of amorphous tungsten silicide has an interesting and strong dependence on stoichiometry. That is, pure tungsten has $T_C$ as low as ~15 mK,[17] while pure silicon is not superconducting. Yet, following earlier work by Kondo on thick $W_xSi_{1-x}$ films, $T_C$ is increased by more than two orders of magnitude at intermediate Si doping values. Based on this work, we aimed at $T_C$ higher than 3 K, which is obtained at $x \approx 0.3$.[18] Some earlier works on $\alpha W_xSi_{1-x}$- based SNSPDs involved co-sputtering of W and Si.[19] To simplify the growth procedure, we used an $\alpha W_5Si_3$ target and optimized the sputtering parameters mainly based on *in-situ* and room-temperature parameters[20] (hereafter, we refer to our films as $W_5Si_3$). We found that growing at room temperature at 2 mTorr and with Ar:N$_2$ 26.5:6 sccm resulted in the desired film properties. Post-deposition analysis confirmed the amorphous nature of the films (*e.g.* Fig. 1c), while the obtained $T_C$ was 3.9 ± 0.15 K for thicker films (≥ 8 nm). For thinner films (5 nm) we measured $R_s$ = 500 Ω/□ and $T_C$ = 3.7 K, which was the lowest temperature available in our material-evaluation cryostat. Typical cooling curves of such films are illustrated in Fig. 1d. To determine the film thickness, we used ellipsometry methods and compared them with both TEM images and the deposition times.[20] We extracted from these measurements that the real and imaginary indices of refraction were $n = 3.7$ and $k = 3.0$. We then used these values to estimate[21] that the absorptance of bottom-illuminated $W_xSi_{1-x}$ patterned thin films was about half of that of NbN films.[22]

The films were grown on SiO$_2$ for the sake of convenience. However, being amorphous, we expected the film properties to be insensitive to the substrate. Indeed, when we grew in the same batches $\alpha W_5Si_3$ films on different substrates (SiO$_2$ and fused silica MgO and/or Al$_2$O$_3$) the obtained films demonstrated comparable properties ($T_C$ and $R_s$).

Once we optimized the growth conditions of NbN on SiO$_2$ and of $\alpha W_5Si_3$ on various substrates, we grew a 2 nm NbN film on SiO$_2$, after letting it cooling down to room temperature in a UHV system, we grew a top layer of 2 nm αW$_5$Si$_3$ (thicknesses were estimated based on sheet resistance and deposition time). To allow an electrically-transparent interface between the two thin films we kept the gases flowing in between the deposition of the two films, in an effort to maintain the NbN stoichiometry.

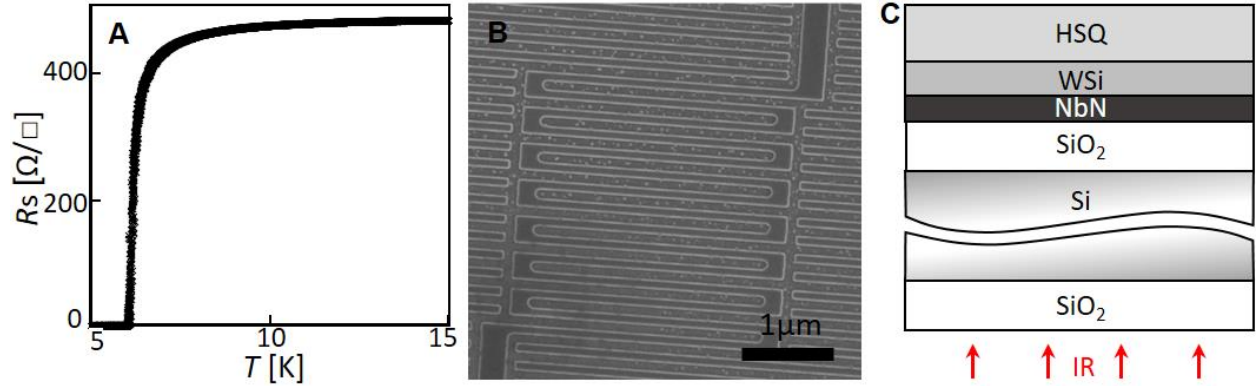

**Figure 2| Hybrid superconducting nanostructures.** (**A**) Cooling curve of a 2-nm-thick α-W$_5$Si$_3$ film on top of a 2-nm-thick crystalline NbN film deposited on SiO$_2$ substrate, demonstrating $T_C = 6.2$ K and $R_s = 447$ Ω/□ of the hybrid film. (**B**) SEM micrograph of an SNSPD made of the hybrid film. (C) Cross-section illustration of the layered structure of the hybrid device. The superconducting bilayer is backlit with IR photons.

The target $R_s$ for the 4 nm bilayer was ~450 Ω/□, which is comparable to our standard NbN-based SNSPD target sheet resistance. Seven growth runs were performed, in which an NbN/αW$_5$Si$_3$ stack was deposited with varying deposition time. In these depositions, we maintained a constant ratio of the deposition times of NbN to αW$_5$Si$_3$, while varying the total time (and thus thickness, estimated to be between 3 and 10 nm). The final wafer used for testing had $R_s = 447$ Ω/□, and $T_C = 6.2$ K after 50-sec deposition of the NbN and 42-sec deposition of the αW$_5$Si$_3$. A cooling curve of the entire stack is given in Fig. 2a, demonstrating a rather smooth (RRR = 0.94) and abrupt ($\Delta T_C = 0.55$ K) superconducting transition. The sister chip from the same growth batch had had the same $T_C$ and $R_s$, within our measurement error, demonstrating the robustness of our growth process. The physical device and schematics of the optical stack are given in Figures 2b-c.

### Results

The SNSPD fabrication procedure of the bilayer NbN/αW$_5$Si$_3$ followed the process reported elsewhere.[20] The room temperature resistance of our 3.2 × 3.2 µm$^2$ devices was: 291.0 kΩ and the critical superconducting current was 9.1 µA at 2.6 K.[20,23] Measuring the time evolution of the voltage on the SNSPD as a result of a single-photon detection event and averaging over 100 such events, we found a typical < 5 ns reset time (Fig. 3a). We examined also the rise time of 25000 photon detection events and determined the characteristic jitter to be 52 ps (Fig. 3b). Lastly, we measured the device detection efficiency as a function of applied current bias on the device and found that the detection efficiency profile was nearly saturated (> 96% of the saturating value).

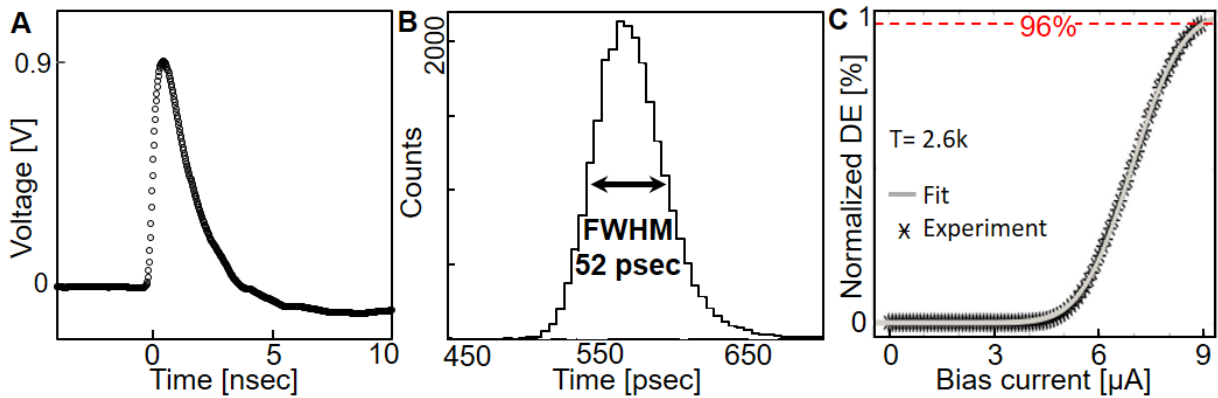

**Figure 3| Hybrid superconducting-nanowire single-photon detector**. (**A**) Averaged of over 100 photo-detection events in the SNSPD demonstrate < 5 ns rest time. (**B**) The full-width half maximum of a histogram of the difference in arrival time of 25000 pulses indicates a 52 ps timing jitter. (**C**) Detection events as a function of the bias current applied on the SNSPD illustrate that upon increasing the bias current, the efficiency profile first increases sharply and then reach

96% of the saturation value near $I_s$, suggesting that the device is operating near its internal quantum-efficiency limit. All measurements were done at 2.6 K.

**Discussion**

The measured characteristics demonstrate indeed a successfully optimized SNSPD that combines several of the advantageous properties of SNSPDs that usually do not coexist. That is, the < 5 ns reset time is comparable to NbN devices and is faster than a typical $W_xSi_{1-x}$ SNSPD, while same applies to the 52 ps timing jitter. The detection-efficiency profile as a function of bias current suggests that the device operates near its internal quantum efficiency limit. We quantified this claim by fitting this profile (Figure 3c) to an error function, which showed that the maximum device efficiency was 96% of the saturated asymptotic value. We confirmed that this profile should fit to this symmetric cumulative distribution function by: (*i*) verifying that the profile of our measurement is antisymmetric; and (*ii*) successfully fitting previously-reported[2] saturating profiles of $W_xSi_{1-x}$ to an error function.[2] In addition to the timing and efficiency performance, our devices demonstrated a switching current of 9.1 µA. This value is higher than typical critical currents of similar geometries measured at similar temperatures for WSi.[2,3,24] Critical current is a key indicator of signal to noise ratio, and thus impacts device jitter and false count rates.

Using standard thin-film optical modeling methods,[21,25] we estimated the optical absorptance analytically based on indices of refraction for NbN.[21] We determined that for the typical fill factors and thickness we used, the absorptance was slightly above the value calculated by simply averaging the absorptance of NbN and WSi films with the same thickness.[22] These properties allowed us to maintain a high $I_C$, adequate absorptance, while achieving nearly-saturated operation.

Our measurements demonstrated that a bilayer NbN- $W_xSi_{1-x}$ SNSPD exhibits optimized and improved properties with respect to single-layer devices. Here, we elaborate why the mechanism leading to this improvement is most likely a result of hybridization of the superconducting electrons of the two materials.

To begin with, single-material devices have fixed intrinsic properties, such as kinetic inductivity, resistivity, $T_C$, critical current density, specific heat, thermal conductivity, and optical properties. The lack of tunability of these properties in such devices limits their performance. Hence, hybridizing materials allows one to tune these properties and the device performance. For instance, single-material-based SNSPDs require a design tradeoff between speed and internal quantum efficiency. On the other hand, a hybrid-material-based device enables control of each of these two characteristics by using different materials with different properties in the stack, allowing optimized performance. In addition to the device properties, the bilayer sheet resistance was measured as the sum of the 2-nm-thick NbN and $W_5Si_3$ films ($1/R_s\text{hybrid}=1/R_s\text{NbN} +1/R_sW_xSi_{1-x}$), indicating that both films were deposited successfully.

We used several independent methods to establish that the hybrid films were superconducting. First, the $I_C$ measured for the device was higher than that expected for a 2nm thick NbN wire of a similar cross section. Moreover, the measured reset time and timing jitter in the bilayer device were faster than a typical $W_xSi_{1-x}$ SNSPD, but slower than a typical NbN SNSPD. In fact, if only the NbN was superconducting in the bilayer SNSPDs, because of the expected increase in kinetic inductivity of the thinner film, these timing properties should have become slower, and not faster than a 4nm thick NbN SNSPD. Furthermore, the saturating profile of the efficiency *vs.* bias current (Figure 3c) was typical to a $W_xSi_{1-x}$ and not typical to an NbN SNSPD at 1550-nm wavelength (recent results for SNSPDs on $Si_3N_4$[26] and AlN[27] suggest that the saturation characteristic is sensitive to the substrate as well as to the nanowire material). In addition, the normal-to-superconducting transition of the bilayer film $\Delta T_C$ (Figure 2b) was narrower than a typical $\Delta T_C$ of a NbN[14] film and closer to the typical $\Delta T_C$ of $W_xSi_{1-x}$ films.[28,29] While we observed a full superconducting transition for the thicker $W_xSi_{1-x}$ films, a 5-nm $W_xSi_{1-x}$ film exhibited only a partial transition as it approached the 3.8 K minimum temperature of our materials-analysis cryostat. Taken together, these observations support the conclusion that at the 2.6 K temperature of our device-characterization cryostat, both films in the hybrid material stack were superconducting.

The mechanism of this hybridization is straightforward: because of the intimate contact of the 2nm-thick $\alpha W_5Si_3$ film to the NbN in the stack, Cooper pairs from the NbN film penetrate the $\alpha W_5Si_3$ (and vice versa) due to the proximity effect. That is, the coherence length of the induced electron pairs satisfies: $\xi=\hbar v_f/(2\pi k_B \cdot T)$ at the clean limit. For the dirty limit: $\xi=(\hbar D/(2\pi k_B \cdot T))^{1/2}$, while $D = v_f \cdot L/3$ is the diffusion coefficient.[30] Here $\hbar$, $v_f$ and $k_B$ are the reduced Planck's constant, Fermi velocity and Boltzmann's constant respectively. Thus, for $\alpha W_5Si_3$ ($D = 0.61\times10^{-4}$ m$^2$/sec) [31] at $T = 2.6$ K, the induced coherence length is $\gtrsim$ 2nm- the film thickness, suggesting that the $\alpha W_5Si_3$ must become a superconductor even if it would have not been such in the first instance.

In fact, the $\alpha W_5Si_3$ film is expected to enjoy dual superconducting characteristics simultaneously. On the one hand, the film should be a 'native' superconductor with Cooper pairs that are typical for such films. On the other hand, there are superconducting electrons that migrated from the nearby NbN film thanks to the proximity effect. These electrons have a coherence length that is larger than the 2nm $W_5Si_3$ film thickness. Thus, within the $\alpha W_5Si_3$ film, the native superconductivity characteristics and the superconducting characteristics of the NbN-proximitizing superconductivity coexist. Likewise, the NbN also should exhibit a dual superconducting characteristic. On the one hand, the as-grown 2-nm-thick film is a native superconductor. On the other hand, because the NbN is in close proximity to the superconducting $\alpha W_5Si_3$ film, superconducting electrons migrate also from the $\alpha W_5Si_3$ to the NbN film thanks to the proximity effect. Substituting $D=0.5\times10^{-4}$ $m^2$/sec for NbN[31] into the coherence length of the proximitized Cooper pairs, we obtain that $\xi$ is larger than the 2nm film thickness. That is, similarly to the $\alpha W_5Si_3$, the entire NbN film exhibits two types of superconducting behavior simultaneously- native superconductivity and proximitized superconductivity.

Because the two films in the bilayer both exhibit simultaneously native superconductivity and proximitized superconductivity that originates from the neighboring film, we can say that the bilayer forms hybrid superconductivity. As a result, the superconducting properties of the granular NbN film and the amorphous $W_5Si_3$ film are hybridized. In Figure 4, we illustrate schematically the hybridization mechanism, which is supported by our experimental observations of a combination of the advantageous SNSPD properties in the bilayer device (Fig. 3).

The proximity effect has been explored and technologically utilized previously.[32–36] Typically, a superconductor was brought in close proximity with a normal metal either to induce superconductivity in the metal or to influence the superconducting properties of the superconductor.[12] In addition, the proximity effect between two superconducting materials of different $T_C$ has been explored.[37] When the ambient temperature is between these two $T_C$ values, superconductivity is induced in the lower $T_C$ material, even though it would have not been a native superconductor at that temperature. These proximity effects are different than the hybrid mechanism we propose. Here, the two materials exhibited native superconductivity, while at the same time they hosted the proximitized superconductivity characteristics.

The coherence length of the proximitized electrons is expected to be only a few nanometers. Thus, the proximity effect is considered an interface effect, where the proximitized electrons exist only around the interface between the superconductor and the other material. In previous studies, either the superconductor or the normal material or both were thicker than this length scale so that the proximitized electrons exist only in a small region of the entire structure. However, in the framework of the current study, each of the films was thinner than the proximitized coherence length. Thus, the entire structure was essentially a single interface so that the proximitized and the native superconductivity coexisted, allowing true hybridization (as illustrated in Fig. 4).

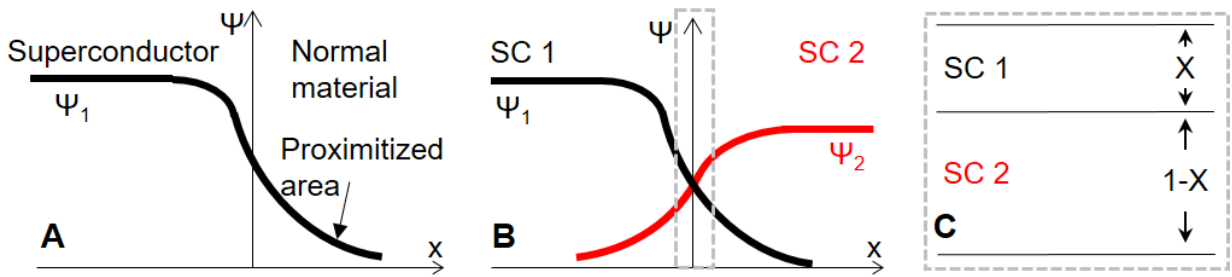

**Figure 4| Schematic illustration of the superconducting-superconducting hybridization mechanism**. (**A**) A superconductor proximitizing an adjunct normal material. (**B**) Two neighboring superconductors proximitize each other at the interface. In (A-B) the proximity effect takes place only at the interface between the two materials, while the bulk properties dominate the behavior further away from that interface. However, in systems such as those discussed here, where the physical size is smaller than the proximitized coherence length (highlighted with a dashed frame in B) both the original and proximitized superconducting behavior coexist throughout the entire material so that the entire structure becomes a hybrid superconductor with a dual-superconducting characteristics. (**C**) The properties of the hybrid can be tuned by adjusting the relative thickness for the superconducting films in the stack, highlighting one superconductor characteristics over the other.

Page 14 of 27

The key result of this paper is the demonstration of a method by which a superconducting thin-film device can be optimized by combining the properties of dissimilar materials in close proximity. That is, traditionally, one has to choose a specific material for optimizing specific properties and compromising on the others (*e.g.*, a device can be either fast or efficient, but not both at the same time). We demonstrated performance of the hybrid device exceeded that of a single-material SNSPD made of NbN or $W_xSi_{1-x}$ with respect to critical current, jitter, and internal quantum efficiency. We should note that based on recent designs of NbTiN SNSPDs with improved performance,[38] hybridization of NbTiN and other materials with $W_xSi_{1-x}$ or superconducting with complimentary properties may enhance the performance even further and should be examined. We also proposed the mechanism that we believe governs the behavior of a hybrid NbN-WSi device. Here we studied SNSPDs as an example, but this technique could also be used to optimize properties of other superconducting-based nanotechnologies (*e.g.,* nTron,[39] yTron,[40] resonators,[41] MKIDs,[42] parametric amplifiers,[43] nanoSQUIDs[44] and Dayem-Bridge devices[45] with applications across information processing, communications, and magnetic and radiation sensing). For such applications, continuous tuning of material parameters may be achieved. In general, one could vary the relative ratios of the materials, or even combine more than two materials to further fine-tune the device characteristics as illustrated in Fig. 4c.

The framework to which the effect of superconducting hybridization is significant exceeds the field of SNSPDs and is not only relevant for nano-superconducting technologies, but also helps understand some of the fundamental behavior of superconductivity near the superconducting-to-insulating transition (SIT).[14,46,47] To date, despite the renewal interest in the vortex-pairing model of Kosterlitz and Thouless,[48] there is no general agreement over a single model that describes the SIT and applies to all superconducting materials.[14,49–54] It is believed that the main distinction between models is between granular and amorphous (*i.e.,* homogeneous) materials.[14,51] Nevertheless, although tuning parameters, such as film thickness, sheet resistance, material stoichiometry and magnetic field are readily available experimentally, the tunability of granularity or amorphousness of superconductors is more challenging, even though it is highly desired. That is, there do not exist convenient methods by which one can study the mechanism of transitions between these two descriptions without affecting other parameters. Therefore, the method suggested here to hybridize amorphous and granular superconductors (here, $W_5Si_3$ and NbN) may also be applied not only to enhance the performance of SNSPDs or superconducting-based technologies in general, but also to help continuously tune the amorphousness of a superconductor. Such a method could be used to elucidate the onset of superconductivity near the SIT (an example of such tunability is given in Fig. 4c).

Acknowledgements

We acknowledge support from the U.S. Office of Naval Research (contract #N00014-13-1-0074) and the DARPA DETECT Program, while AED was supported by the iQuISE fellowship. YI and MB thank also the Horev Fellowship for Leadership in Science and Technology, supported by the Taub Foundation, the Zuckerman STEM Leadership Program as well as the Technion's Russel Berrie Nanoscience Institute and Grand Energy Program.

Authors' Contribution

YI and KKB initiated the project. YI grew the NbN films, while JJS grew the WSi films and performed the electrical characterisation. CSK performed the TEM imaging with the help of YI and EKC. FN fabricated the patterned devices. YI and JJS characterised the devices, with the support of YI, KKB, JJS, AED and MB analysed the data. YI, KKB and MB wrote the paper. All authors help proofread the paper.

Supplementary Information

## 1. Fitting of reset time

The voltage waveform produced when an SNSPD responds to an incident photon can be used to estimate the kinetic inductance of the nanowire, as long as the readout circuit is properly accounted for. The SNSPD voltage waveform as a function of time has a typical shape that can be divided into two regimes: (1) a sharp initial rise caused by the rapid appearance of a large resistance in series with the nanowire which causes the bias current to divert to the readout, (2) a slow decay as the bias current returns to the nanowire. The decay process is usually dominated by the kinetic inductance of the nanowire ($L_k$), and if the readout circuit is a pure resistance with value R, then the SNSPD reset time can be fitted to a falling exponential function with time constant of $L_k/R$. However, in our case, the SNSPD was connected to readout amplifiers via the RF port of a bias T as shown below. If we assume that the lower cutoff of the amplifier is a simple *RC* high-pass filter with f3db of 20 MHz, we can estimate the amplifier input capacitance to be approximately 160 pF. In series with the 200 nF capacitance of the bias T, the amplifier capacitance will dominate, leading to a readout circuit that is well approximated by adding a 160 pF capacitor in series with a 50-Ω resistor. Solutions to the underdamped *RLC* circuit can then be matched to our experimental results in order to estimate the kinetic inductance of the nanowire (Fig. SI1).

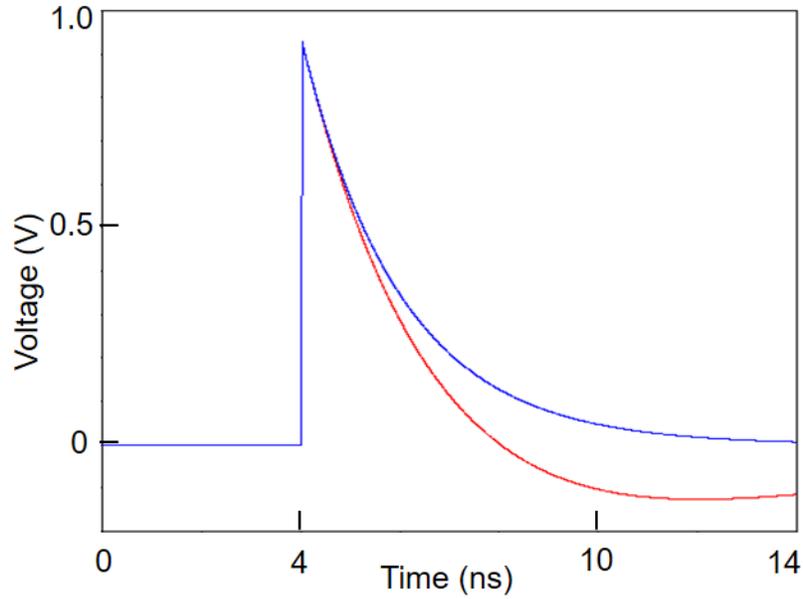

**Figure SI1|** A simulated SNSPD pulse is shown with a 200 nF capacitor (blue) and a 160 pF capacitor (red).

## 2. Growth

Prior to processing the hybrid film, we optimized the growth conditions independently for both NbN and $\alpha W_5Si_3$. We aimed at a hybrid film of a 4-nm thickness, similarly to those on which we reported in several of our previous works.[9,14,21,23,25,26,55] We found that the resistivity of the two materials is rather close for the films with thicknesses in the desired range. To demonstrate the full effect of the hybridization, the device we were interested to process had to comprise an equal volume for the two materials, *i.e.*, the hybrid film should have been made of 2-nm thick NbN and $\alpha W_5Si_3$ films. Measuring accurately the thickness of such ultra-thin films that are made of different chemical compounds is a complex task.[14] However, measuring the sheet resistance is much more realistic. Thus, we chose to aim at a hybrid stack, which includes NbN and $\alpha W_5Si_3$ films, while for each of them $R_s \approx 1000\ \Omega/\square$ (ca. 2 nm). The growth conditions of the NbN films are given elsewhere,[14] with the only difference that we chose a Si wafer, with 255-nm thick thermalized SiO$_2$ film on both sides. Because Si is opaque, the actual substrate temperature is expected to be higher than that of the MgO substrate, on which we grew the NbN earlier, although the nominal heater temperature did not change. In practice, the NbN film was deposited by means of reactive DC magnetron sputtering, where we used an Nb target and introduced Ar and N$_2$ in the chamber with 26.5, and 6 sccm, respectively, while the total pressure in the chamber was 2.5 mTorr. Based on our previous experience, we deposited the NbN for 50 sec to obtain the desired thickness (or sheet resistance). All superconducting films were sputtered with AJA Orion.

To grow the $\alpha$W$_5$Si$_3$, we used a single target with the desired stoichiometry. We first examined our films' quality by comparing them to the work of Kondo,[18] in which he studied superconductivity in $\alpha$W$_x$Si$_{1-x}$ films. Specifically, we measured $T_c$ = 3.9 K for films as thin as 5 nm ($R_s$ = 239.7 Ω/□), which is in a good agreement with the $T_c$ = 5 K of Kondu's films that were as thick as 500 nm. Our films were grown under the optimized conditions that include 26.5 sccm, while the total pressure was 2 mTorr. We found that the desired sheet resistance is obtained by depositing the films for 42 sec.

To maintain the amorphousness of the W$_5$Si$_3$, we grew these films at room temperature. Moreover, given the significance of the interface between the two materials, we sputtered the entire hybrid film without removing the sample from the sputtering chamber, which had a < 5·10$^{-9}$ mTorr base pressure. Hence, prior growing the $\alpha$W$_5$Si$_3$ on the NbN film, we waited for the chamber to equilibrate at room temperature. To maintain the stoichiometry of the NbN film close to the interface, we maintained the Ar and N$_2$ at the same conditions as they were flowing during the NbN growth until the nominal temperature of the chamber decreased to 100°C, below which there were no gases flowing in the chamber and the vacuum returned to its based pressure.

The post-growth film characterization follows the procedure we reported elsewhere.[14] Likewise, both the device preparation and characterization procedures are reported elsewhere.[3]

### 3. Optical performance

In addition to the quantum efficiency, we can compare the calculated optical absorption of the different materials—NbN, αW$_5$Si$_3$ and the hybrid structure in absolute numbers for the detected 1550-nm-wavelength photons. Such a calculation requires the real and imaginary (*n* and *k*) parts of the index of refraction of these materials. We recently reported a thorough study of the NbN's index of refraction,[21] while for the αW$_5$Si$_3$ we performed ellipsometry characterizations. By fitting the measured *n* and *k* in several wavelengths to Cauchy's equation, we extracted $n_{1550nm}$ = 3.7, $k_{1550nm}$ = 3.0 (Fig SI2).

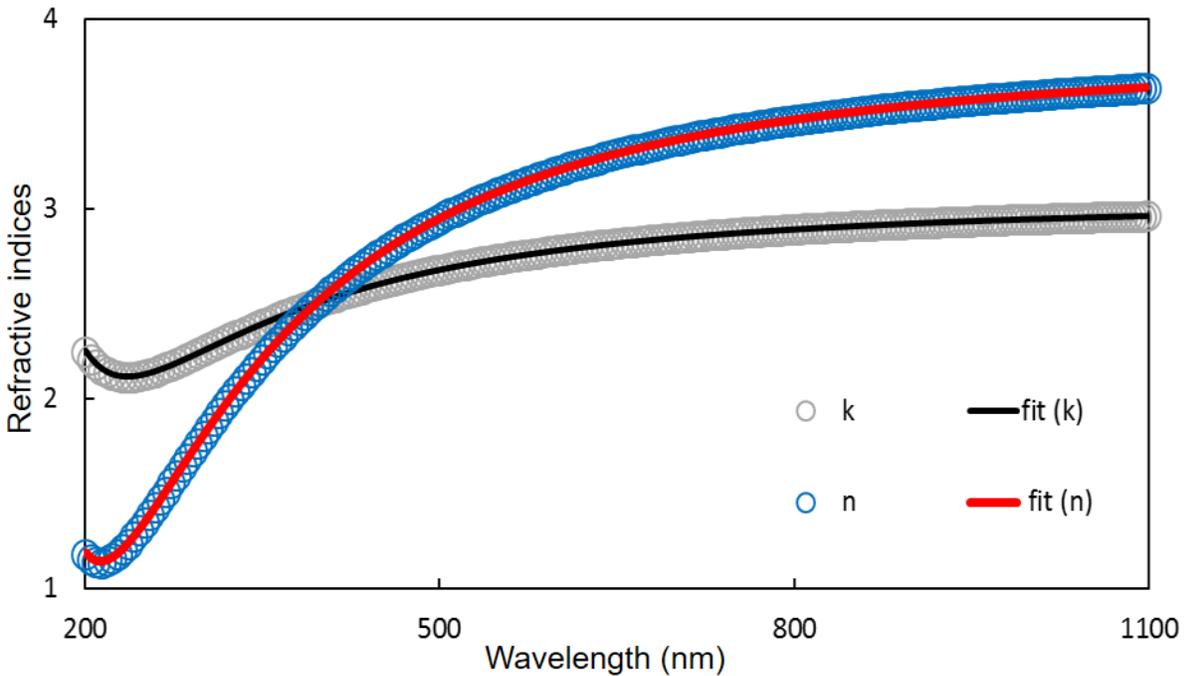

**Figure SI2| Ellipsometry characterizations of αW$_5$Si$_3$**. The refractive indices (*n* and *k*) are marked with circles, while the best fit to Cauchy equation is the full lines.

Moreover, given that unlike a continuous film, an SNSPD is a meander structure that covers only a certain fraction of the area, the fill-factor of an SNSPD is needed to estimate the actual device absorption. Figure SI3 illustrates a comparison between the absorption of 4-nm-thick SNSPDs made of these materials for a broad range of fill factors. According to these calculations, we found that the quantum efficiency of our hybrid device is 15-25%.

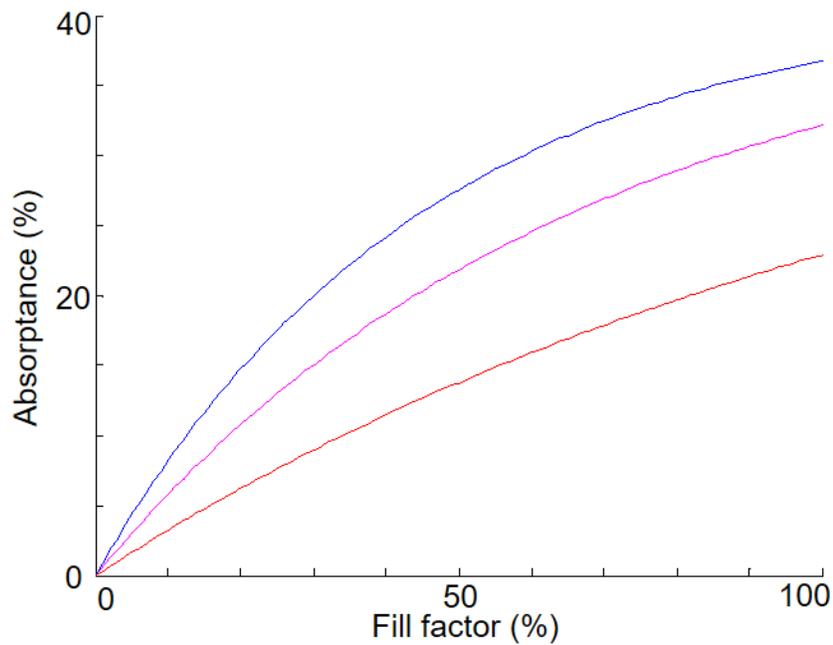

**Figure SI3| Calculated absorptance versus fill factor** for 4-nm-thick NbN (blue), 4-nm-thick $\alpha W_5Si_3$ (red) and a bilayer of 2-nm-thick NbN under 2-nm-thick $\alpha W_5Si_3$ (purple).

We should mention that for practical applications, by integrating cavities on the devices, this absorption can be doubled,[28] while by introducing antennae, the efficiency can be doubled again. Therefore, our demonstration of saturating device (Figure 3) suggests that the hybrid devices can operate at system efficiency near unity with the high timing performance we reported here, and at 2.6 K. That is, bringing SNSPDs close to their ultimate performance.